\documentclass[useAMS,usenatbib]{mn2e}
\usepackage{amsmath,array,amsfonts}
\usepackage{amssymb}
\usepackage{graphicx}
\usepackage{natbib}
\usepackage{pstricks}
\usepackage{verbatim}
\usepackage{multirow}
\usepackage{color, colortbl}
\usepackage{subfigure}
\usepackage{booktabs}
\definecolor{Gray}{gray}{0.7}

\newcommand{\ltsima} {$\; \buildrel < \over \sim \;$}
\newcommand{\gtsima} {$\; \buildrel > \over \sim \;$}
\newcommand{\lta} {\lower.5ex\hbox{\ltsima}}
\newcommand{\gta} {\lower.5ex\hbox{\gtsima}}

\title[Local $f_{\rm NL}$ estimators]
{Exploring local $f_{\rm NL}$ estimators based on the binned bispectrum}
\author[B. Casaponsa et al.]{B. Casaponsa$^1$ $^2$\thanks{e-mail:
casaponsa@ifca.unican.es}, R.B. Barreiro$^1$, E. Mart\'{i}nez-Gonz\'{a}lez$^{1}$, A. Curto$^{1,3}$, M. Bridges$^3$,\newauthor M.P. Hobson$^3$\\ 
$^1$     Instituto de F\'isica de Cantabria, CSIC-Universidad de Cantabria, Avda. de los Castros s/n, 39005 Santander, Spain.\\
$^2$     Dpto. de F\'isica Moderna, Universidad de Cantabria, Avda. los Castros s/n, 39005 Santander, Spain.\\
$^3$ Astrophysics Group, Cavendish Laboratory, J.J. Thomson Avenue, Cambridge CB3 0HE, U.K.}
\date{Accepted ---. Received ---; in original form ---}
\begin{document} 
\maketitle

\begin{abstract}  
We explore different estimators of the local non-linear coupling
parameter, $f_{\rm NL}$, based on the binned bispectrum presented in
Bucher et al.  Using simulations of Wilkinson Microwave Anisotropy
Probe (WMAP)-7yr data, we compare the performance of a regression
neural network with a $\chi^2$-minimization and study the dependence of
the results on the presence of the linear term in the analysis and on
the use of inpainting for masked regions. Both methods obtain similar
results and are robust to the use of inpainting, but the neural
network estimator converges considerably faster.  We also examine the
performance of a simplified $\chi^2$ estimator that assumes a diagonal
matrix and has the linear term subtracted, which considerably reduces the
computational time; in this case inpainting is found to
be crucial. The estimators are also applied to real WMAP-7yr data,
yielding constraints at 95\% confidence level of $-3<f_{\rm NL}<83$.
\end{abstract} 

\section{Introduction}

Cosmic microwave background (CMB) fluctuations naturally arise in
inflationary models. Discriminating between different models is a
difficult task, but can be addressed by observing very faint
non-Gaussian signals in the high-order correlation functions of the
CMB temperature anisotropies.  A popular approach is to search for the
local form of non-Gaussianity, where the initial curvature Gaussian
perturbations are expanded up to the second order
as \[\Phi=\Phi_g+f_{\rm
  NL}\left[\Phi^2_g-\left<\Phi^2_g\right>\right]\] \citep[for more
  details see e.g.][]{Bartolo2004,Babich2004}.

WMAP constraints on the amplitude of the local form of non-Gaussianity
have been able to rule out exotic models such as ghost inflation
\citep{Arkani2004}. New data sets, such as the recent release from
Planck satellite \citep{Planck2013}, significantly reduce the
uncertainties on local $f_{\rm NL}$, ruling out the ekpyrotic model
and imposing strong constraints on multi-field inflationary models.
In fact, for single-field inflation, $f_{\rm NL}$ (hereafter $f_{\rm
  NL}$ is the local form) should be of the order of the spectral index
\citep{Creminelli2004}, given the consistency relation derived in
\cite{Maldacena2003}. Recent papers show that this relation does not
hold for non-vacuum initial states \citep{Ganc2011,Agullo2011} and
non-constant super-horizon modes \citep{Chen2013}, but the vast
majority of single-field models should be ruled out by a detection of
a larger $f_{\rm NL}$ value.

This type of primordial non-Gaussianity may be detected using
higher-order correlation functions. The simplest of these is
third-order, which is equivalent to the bispectrum in spherical
harmonic space. The first derivation of the optimal estimator, in the
sense of an unbiased estimator that saturates the Cramer--Rao
inequality, is given in \citet{Babich2005}, assuming an isotropic
field.  Working with real data, however, is usually more
complicated. In particular, CMB maps have anisotropic noise due to the
scanning strategy and masked regions, both of which break the isotropy
assumption for these theoretical estimators. The masked regions are
particularly difficult to treat, as they introduce correlations among
the Fourier modes, which are otherwise expected to be
independent. \citet{Creminelli2006} applied the optimal estimator to
real data, showing that the presence of a term proportional to the
$a_{\ell m}$ is required to account for such anisotropies. In that
paper the constraints are computed using an approximation to avoid
numerical difficulties. Finally, this estimator was successfully
applied in its complete form to WMAP data by \citep[][for 5th, 7th and
  9th year respectively]{Smith2009,Komatsu2011,Bennett2012}.

New imaging reconstruction techniques have recently been used to
pre-process CMB maps by smoothing the contours of the masked
regions. A simple approach is to apodize the mask by introducing a
smooth function in the pixels surrounding the masked regions. Another
approach is to fill the masked regions with a pseudo-signal, which is
termed inpainting. Several techniques have been proposed in the
literature for inpainting, which is a very delicate process since the
signal can be distorted \citep{Bajkova2005,Abrial2008,Starck2013}.

Consequently, primordial non-Gaussianity analyses can be
computationally demanding, and new techniques should therefore be
investigated to overcome the computational cost of large matrix
estimations and inversions. Here we investigate the utility of a
neural network to obtain the necessary weights in the $f_{\rm NL}$
estimator and compare it with the direct approach via $\chi^2$
minimization. Over the last 20 years, artificial intelligence
techniques have been use in a number of areas of astrophysical
analysis: morphological galaxy determination, photo-redshift
estimations, and classification of different objects are examples of
successful applications of neural networks \citep{Storri-Lombardi1992,Firth2003,Vanzella2004,Carballo2008}.  In
particular, for cosmological analysis, they have recently been used to
reduce the computational time of cosmological parameter estimation
from observations of the CMB power spectrum
\citep{Auld2007,Auld2008}. Also in CMB analysis,
\citet{Casaponsa2011b} used neural networks to define a new
non-Gaussianity estimator and showed that networks are a valuable tool
for bypassing the inversion of ill-conditioned matrices, and to avoid
covariance matrix estimation in a $\chi^2$ analysis. 

The aim of the present work is to continue our earlier study of the
power of the neural networks in the statistical analysis associated
with cosmic microwave background (CMB) non-Gaussianity.  To this end,
this paper is focused on the study of different tools, in order to
identify the most robust and efficient estimator when dealing with
real data. We compare three different approaches to estimate $f_{\rm
  NL}$, based on the binned bispectrum. The first estimator is
obtained by minimizing a $\chi^2$ of the binned
bispectrum components. A second approach is based on the optimal
estimator, without taking into account the correlations among the
binned bispectrum components, which for a isotropic field would be the
same as the former. And the third method uses the weights of a
regression neural network. From these approaches we construct
different estimators to account for the effects of pre-processing the
data with inpainting and the presence of a the linear term.

The paper is organized as follows. An overview of the type of neural
network employed and the training procedure is given in
Section~\ref{sec:nn}. In Section~\ref{sec:binned_bispectrum} we
describe the binned bispectrum. The definition of the estimators is
presented in Section~\ref{sec:estimators} followed by an explanation
of the main details of the implementation in
Section~\ref{sec:method}. The results are presented in
section~\ref{sec:results}, and finally the conclusions are summarised
in section~\ref{sec:conclusions}.

\section{Neural networks}
\label{sec:nn}

Artificial neural networks (ANN) are a methodology for computing,
based on massive parallelism and redundancy, which are features also
found in animal brains. They consist of a number of interconnected
nodes each of which processes information and passes it to other nodes
in the network. Well-designed networks are able to `learn' from a set
of training data and to make predictions when presented with new,
possibly incomplete, data.  These algorithms have been successfully
applied in several areas, in particular, we note the following
applications in cosmology: \cite{Baccigalupi2000,Firth2003,Ball2004,Auld2007,Auld2008,Casaponsa2011b}
and \cite{Norgaard2012}.
 
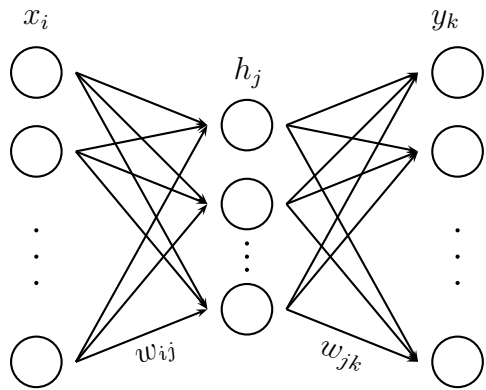
\begin{figure}
 \center
\psset{unit=0.7cm}
\begin{pspicture}(0,0)(10,8)
\uput[0](0.5,7){\Large$x_{i}$}
\pscircle(1,6){0.5}
\pscircle(1,4.5){0.5}
\pscircle(1,3){0.001}
\pscircle(1,2.5){0.001}

\pscircle(1,2){0.001}
\pscircle(1,0.5){0.5}
\rput[bl]{20}(3,0.3){\Large$w_{ij}$}
\psline{->}(1.75,6)(4.25,5)
\psline{->}(1.75,6)(4.25,3.5)
\psline{->}(1.75,6)(4.25,1.5)
\psline{->}(1.75,4.5)(4.25,5)
\psline{->}(1.75,4.5)(4.25,3.5)
\psline{->}(1.75,4.5)(4.25,1.5)
\psline{->}(1.75,0.5)(4.25,5)
\psline{->}(1.75,0.5)(4.25,3.5)
\psline{->}(1.75,0.5)(4.25,1.5)
\uput[0](4.5,6){\Large$h_{j}$}
\pscircle(5,5){0.5}
\pscircle(5,3.5){0.5}
\pscircle(5,2.75){0.001}
\pscircle(5,2.5){0.001}
\pscircle(5,2.25){0.001}

\pscircle(5,1.5){0.5}
\rput[bl]{-18}(6.3,0.5){\Large$w_{jk}$}
\psline{->}(5.75,5)(8.25,6)
\psline{->}(5.75,5)(8.25,4.5)
\psline{->}(5.75,5)(8.25,0.5)
\psline{->}(5.75,3.5)(8.25,6)
\psline{->}(5.75,3.5)(8.25,4.5)
\psline{->}(5.75,3.5)(8.25,0.5)
\psline{->}(5.75,1.5)(8.25,6)
\psline{->}(5.75,1.5)(8.25,4.5)
\psline{->}(5.75,1.5)(8.25,0.5)
\uput[0](8.25,7){\Large$y_{k}$}
\pscircle(9,6){0.5}
\pscircle(9,4.5){0.5}
\pscircle(9,3){0.001}
\pscircle(9,2.5){0.001}
\pscircle(9,2){0.001}
\pscircle(9,0.5){0.5}

\end{pspicture}
\caption{Schematic diagram of a 3-layer feed-forward neural network.}
\label{fig:network_diagram}
 \end{figure}

The basic building block of an ANN is the \emph{neuron} or {\em
  node}. Information is passed as inputs to the neuron, which
processes them and produces an output.  The output is typically a
simple mathematical function of the inputs. The power of the ANN comes
from assembling many neurons into a network. The network is able to
model very complex behaviour from input to output. We use a
three-layer feed-forward network consisting of a layer of input
neurons, a layer of `hidden' neurons and a layer of output neurons.
Figure~\ref{fig:network_diagram} shows a schematic design of such a
network.

The outputs of the hidden layer and the output layer are related to
their inputs as follows:
\begin{eqnarray}
\mbox{hidden layer:} & h_j=g^{(1)}(f_j^{(1)}); &
f_j^{(1)} = \sum_i w^{(1)}_{ji}x_i +
  \theta_j^{(1)}, \\
\mbox{output layer:} & y_k=g^{(2)}(f_k^{(2)}); & f_k^{(2)} =
\sum_j w^{(2)}_{kj}h_j + \theta_k^{(2)},\label{eq:outputs}
\end{eqnarray}
for each hidden node $j$ and each output node $k$. The index $i$
runs over all input nodes. The functions $g^{(1)}$ and $g^{(2)}$ are
called activation functions. The non-linear nature of $g^{(1)}$ is a
key ingredient in constructing a viable and practically useful
network. This non-linear function must be bounded, smooth and
monotonic; we use $g^{(1)}(x) = \tanh x$. For $g^{(2)}$ we simply use
$g^{(2)}(x) = x$.  The layout and number of nodes are collectively
termed the \emph{architecture} of the network. For a basic introduction 
to artificial neural networks the reader is directed to \citet{MacKay} and \citet{Golden1996}.

For a given architecture, the weights $\mathbf{w}$ and biases
$\mathbf{\theta}$ define the operation of the network and are the
quantities we wish to determine by some \emph{training}
algorithm. Basically, the training process is an iterative algorithm
that optimises a given objective function that quantifies the accuracy
of the network outputs. We denote $\mathbf{w}$ and $\mathbf{\theta}$
collectively by the network parameters $\mathbf{a}$. As these
parameters vary during training, a very wide range of non-linear
mappings between inputs and outputs is possible. In fact, according to
a `universal approximation theorem'~\citep{Leshno1993}, a standard
three-layer feed-forward network can approximate any continuous
function to \emph{any} degree of accuracy with appropriately chosen
activation functions and a sufficient number of hidden nodes.
 
In our previous application of ANN to the estimation of $f_{\rm NL}$,
a classification neural network was used \citep{Casaponsa2011b}. Here,
we instead use a {\em regression} network, which we find to be as
useful as the classification approach, and also allows a
more direct comparison with the $\chi^2$ minimization
procedure. Additionally, using a regression network has the advantage
of reducing the network parameter space, making the training 
faster. 

In \cite{Casaponsa2011b}, we used neural networks for which the inputs
were third-order moments of two wavelet decompositions of the CMB map:
the Healpix wavelet (HW) \citep{Casaponsa2011a} and the spherical
Mexican hat wavelet (SMHW) \citep{Curto2009,Curto2011}. We found the
resulting $f_{\rm NL}$ estimator had the same accuracy as the standard
one based on $\chi^2$-minimization, but was much faster to evaluate.
Here, the inputs to our neural networks are the estimator for the
bispectrum proposed by \citet{Bucher2010}, defined in a number of bins
in $l$-space, which reduces the dimension of the problem by a factor
of $10^5$. 
Our aim is to learn a mapping from the binned bispectrum components of
the (possibly) non-Gaussian CMB (assembled into an input feature
vector $\mathbf{x}$) to the corresponding $f_{\rm NL}$ of the map;
this is discussed in more detail below.  

A suitable objective function for this problem is
\begin{equation}
\mathcal{L}(\mathbf{a}) = \frac{1}{2}\sum_{n}\sum_{k}
[t^{(n)}_k -y^{(n)}_k(\mathbf{x}^{(n)}, \mathbf{a})]^2,
\end{equation}
where the index $n$ runs over the training data-set $\mathcal{D} =
\{\mathbf{x}^{(n)},\mathbf{t}^{(n)}\}$, in which the target vector
$\mathbf{t}^{(n)}$ for the network outputs are the $f_{\rm NL}$
values, as explained in the next section.  One then wishes to find
network parameters $\mathbf{a}$ that minimise this objective function
as the training progresses. This is, however, a highly non-linear,
multi-modal function in many dimensions whose optimisation poses a
non-trivial problem. We perform this optimisation using the {\sc
  MemSys} package \citep{Gull1999}. This algorithm considers the
parameters $\mathbf{a}$ to have prior probabilities proportional to
$e^{\alpha S(\mathbf{a})}$, where $S(\mathbf{a})$ is the
positive-negative entropy functional \citep{Hobson1998}, and $\alpha$
is a hyper-parameter of the prior that sets the scale on which
variations in $\mathbf{a}$ are expected. The value of $\alpha$ is
chosen to maximise its marginal posterior probability, value of which
is inversely proportional to the standard deviation of the prior.
Thus for a given $\alpha$, the log-posterior probability is
proportional to $\mathcal{L}(\mathbf{a}) + \alpha S(\mathbf{a})$. For
each chosen $\alpha$ there is a solution $\hat{\mathbf{a}}$ that
maximises the posterior.  As $\alpha$ varies, the set of solutions
$\hat{\mathbf{a}}$ is called the \emph{maximum-entropy trajectory}. We
wish to find the solution for which $\mathcal{L}$ is minimised which
occurs at the end of the trajectory where $\alpha=0$.  For practical
purposes we start at a large value of $\alpha$ and iterate downwards
until $\alpha$ is sufficiently small so that the posterior is
dominated by the $\mathcal{L}$ term. {\sc MemSys} performs this
algorithm using conjugate gradient descent at each step to converge to
the maximum-entropy trajectory. The required matrix of second
derivatives of $\mathcal{L}$ is approximated using vector routines
only, thus circumventing the need for $O(N^3)$ operations required for
exact calculations. The application of {\sc MemSys} to the problem of
network training allows for the fast efficient training of relatively
large network structures on large data sets that would otherwise be
difficult to perform in a reasonable time. Moreover the {\sc MemSys}
package also computes the Bayesian evidence for the model
(i.e. network) under consideration, \cite[see for example][for a
  review]{Jaynes2003}, which provides a powerful model selection
tool. In principle, values of the evidence computed for each possible
architecture of the network (and training data) provide a mechanism to
select the most appropriate architecture, which is simply the one that
maximises the evidence.


\section{Binned bispectrum}
\label{sec:binned_bispectrum}
Several approaches to bispectrum analyses have been proposed to reduce
the dimensionality of the problem without losing significant
information \citep[see for example][]{Bucher2010,Fergusson2011}. In
particular, we use the bispectrum estimator defined in
\citet{Bucher2010}.  The proposed method consists of joining the
bispectrum components in bins, significantly reducing the
computational time, but maintaining the quality of the estimator of
$f_{\rm NL}$. \citet{Bucher2010} show that this is the case for ideal
maps, with isotropic noise and small symmetric masks. The binned
bispectrum is also applied to Planck data in \citet{Planck2013} to
constrain primordial non-Gaussianity. Here we study with more detail
its applications to realistic data, for which we used simulations 
with WMAP-7yr characteristics.  

As a starting point, the angle-averaged reduced bispectrum is defined
by
\begin{equation} 
 b_{l_{1}l_{2}l_{3}}=\int T_{\ell_{1}}T_{\ell_{2}}T_{\ell_{3}}d\Omega\; ,
\label{eq:bispectrum}
\end{equation}
where $T_{\ell}(\vec{n})=\sum_m a_{\ell m}{\it Y}(\vec{n})$.
 The binned reduced bispectrum is then
\begin{equation}
 b_{abc}
=\sum_{\ell_{1}\in I_{a}}\sum_{\ell_{2}\in I_{b}}\sum_{\ell_{3}\in I_{c}}b_{\ell_{1}\ell_{2}\ell_{3}},
\end{equation}
where $I_{n}$ are bins in $\ell$. This definition of the reduced
bispectrum, differing from the standard one by the factor
$I^2_{\ell_{1}\ell_{2}\ell_{3}}$ \citep[for details
  see][]{Bucher2010,Komatsu2002}, is convenient since one can write
$b_{abc}$ in terms of $T_{a}$, $T_{b}$ and $T_{c}$ which are the
binned maps: 
\begin{equation}
T_{n}=\sum_{\ell_{i}\in I_{n}}T_{\ell_i}. 
\end{equation}
The advantage of constructing maps in $\ell$-bins is that the number
of transformations to spherical harmonic space is significantly
reduced. Then, the resulting bispectrum estimator is faster to
construct than the one based on the KSW estimator \citep{Komatsu2005}
or the SMHW \citep{Curto2011}.

\section{$f_{\rm NL}$ estimators}
\label{sec:estimators}
The optimal estimator for $f_{\rm NL}$, in the sense of an unbiased
estimator that saturates the Cramer-Rao inequality, is obtained by
performing an Edgeworth expansion of the probability distribution of
the $a_{\ell m}$ for weakly non-Gaussian data
\citep{Babich2005,Creminelli2006,Smith2009}. This estimator is found
to have a cubic term and a linear term in $a_{\ell m}$. The latter
term plays an important role under realistic conditions, where
anisotropic instrumental noise and/or a mask is present.

The form of this estimator can also be understood using the properties
of the Wick product.  As demonstrated in \citet{Donzelli2012},
\citet{Marinucci2011} and \citet{Peccati2011}, the Wick product of a
cubic variable, which is given by
\begin{equation} 
:x_1,x_2,x_3: = x_1x_2x_3-x_1\left<x_2x_3\right>-x_2\left<x_1x_3\right>-x_3\left<x_1x_2\right>\;,
\end{equation}
has a smaller variance than the cubic variable itself, while not
affecting the mean value so long as the variables $x_i$ are Gaussian
and have a mean value of zero. Then, if we replace each cubic term in
an estimator by its Wick product, it will yield an estimator with
lower variance. Following this reasoning, the binned bispectrum
defined in Sec.~\ref{sec:binned_bispectrum} can be replaced by its
Wick product
\begin{eqnarray}
:T_{I_{\rm a}}T_{I_{\rm b}}T_{I_{\rm c}}: & = & T_{I_{\rm a}}T_{I_{\rm b}}T_{I_{\rm c}} -\left<T_{I_{\rm a}}T_{I_{\rm b}}\right>T_{I_{\rm c}} \nonumber\\
 &&-\left<T_{I_{\rm b}}T_{I_{\rm c}}\right>T_{I_{\rm a}}-\left<T_{I_{\rm a}}T_{I_{\rm c}}\right>T_{I_{\rm b}}\;.\label{eq:linterm} 
\end{eqnarray}
Note that $T_i=T_i(x)$, since there is a dependence on the pixel for
anisotropic maps.

\citet{Donzelli2012} have proved that for the case of wavelet and
needlet coefficients, the linear term is basically equivalent to removing the
mean value of the coefficients. In order to see if this is the
case for the binned bispectrum, we explore the option of substituting
$T'_n=T_n-\langle T_n\rangle$, where $\langle T_n\rangle$ is computed
with the unmasked pixels. This would be less costly than estimating
the correlation matrix $\langle T_aT_b\rangle$ required for the
computation of the linear term.  

In the following subsections, we describe three methods for choosing
the weights that are used to construct the final $f_{\rm NL}$
estimator. In each case, estimators are constructed with and without
the linear term contribution to explore its importance. Also, the
performance of these estimators is tested on inpainted and
non-inpainted maps, with the methodology explained in
Sec.~\ref{subsec:Inpainting}. In all cases the original mask $M$ is
applied again at the final stage when computing the binned bispectrum
components
\begin{equation}b_{abc}=\displaystyle\sum_{i=1}^{N_{\rm pix}}\frac{M_i(T_{a,i}T_{b,i}T_{c,i})}{4\pi
  N_{\rm pix}},
\end{equation}
where $N_{\rm pix}=\sum_iM_i$.
The efficiency achieved by the estimators will be compared to that
defined by the Cramer-Rao inequality. The Cramer-Rao bound states that the minimum variance for any unbiased estimator is given by the inverse of the Fisher matrix information. A
useful reference value in the case of partial sky coverage is obtained
from the full sky estimator corrected by the fraction of the available
sky. Therefore, the minimum variance for $f_{\rm NL}$ is estimated to
be:
\begin{equation}
\sigma_{fh}^2=\Big[f_{sky}\sum_{\ell_1\le\ell_2\le\ell_3}\frac{\left({\langle B_{\ell_1\ell_2\ell_3}\rangle ^{1}}\right)^2}{\Delta C_{\ell_1}C_{\ell_2}C_{\ell_3}}\Big]^{-1}\label{eq:optimal_sigma}
\end{equation}
where $\Delta$ takes values 1, 2 or 6 when all $\ell$’s are different,
two are equal, or all are the same and $f_{sky}$ is the fraction of
the sky available. For (\ref{eq:optimal_sigma}) to be used for a
realistic case, the power spectrum must include the noise and the beam
contribution. The beam also needs to be included in the bispectrum
part.  We have used WMAP-7yr characteristics, in particular the
average of the two channels of 61 and 94 GHz (V and W) and the
extended mask KQ75. In terms of the reduced bispectrum defined in
Sec.~\ref{sec:binned_bispectrum}, the angular average bispectrum
$B_{\ell_1\ell_2\ell_3}$
is: \begin{eqnarray}B_{\ell_1\ell_2\ell_3}=\sqrt{\frac{4\pi}{(2\ell_1+1)(2\ell_2+1)(2\ell_3+1)}}\times\\\nonumber \left(\begin{array}{ccc}
    \ell_1 & \ell_2 & \ell_3 \\ 0 & 0 & 0 \\
 \end{array} \right)^{-1}b_{\ell_1\ell_2\ell_2}\;.\end{eqnarray}

\subsection{Approximated maximum-likelihood estimator (AMLE)}

The standard approach in this type of analysis is to use the fact that
the third-order moments are nearly Gaussian, and therefore
the maximum-likelihood estimator is obtained approximately by the minimization of a $\chi^2$ given by
\begin{equation}
\chi^2=\sum_{\scriptstyle{abc,def}}\left(b_{abc}-f_{\rm NL}\langle b_{abc}\rangle^{1}\right)C^{-1}_{abc,def}\left(b_{def}-f_{\rm NL}\langle b_{def}\rangle ^{1}\right)\;.\label{eq:chi2}
\end{equation} 
where $\langle b_{def}\rangle^{1}$ is the expected value for $f_{\rm NL}=1$ and $C^{-1}_{abc,def}=\langle b_{abc}\rangle\langle b_{def}\rangle-\langle b_{abc}b_{def}\rangle$. From the previous equation is straightforward to show that the $f_{\rm NL}$ estimator for an observed map is:
\begin{equation} 
f_{\rm NL}=\sum_{\tiny{abc,def}}\frac{\langle b_{abc}\rangle^{1}C^{-1}_{abc,def}b_{def}^{obs}}{\displaystyle\sum_{\tiny{abc,def}}\langle b_{abc}\rangle^{1}C^{-1}_{abc,def}\langle b_{def}\rangle^{1}}\;.
\label{eq:chi2_fnl}
\end{equation}   
In order to include the linear term correction, $T_aT_bT_c$ should be
substituted by its Wick product (\ref{eq:linterm}), wherever it
appears. The expected value of the linear term is zero, and thus it
vanishes in the term of the estimator related to the model, whereas it
needs to be included in the covariance matrix.  Thus, the corresponding
estimator is
\begin{eqnarray}  
f_{\rm NL}=\sum_{\tiny{abc,def}}\frac{\langle b_{abc}\rangle^{1}C^{-1}_{abc,def}}{\displaystyle\sum_{\tiny{abc,def}}\langle b_{abc}\rangle^{1}C^{-1}_{abc,def}\langle b_{def}\rangle^{1}}\times\\\nonumber\Big(\frac{1}{4\pi N_{pix}}\sum_{i}^{N_{pix}}T_{d,i}T_{e,i}T_{f,i}^{obs}\\
\nonumber-\langle T_{d,i}T_{e,i}\rangle T_{f,i}^{obs}-\langle T_{d,i}T_{f,i}\rangle T_{e,i}^{obs}-\langle T_{e,i}T_{f,i}\rangle T_{d,i}^{obs} \Big)\;,
\label{eq:chi2_fnl_lt}
\end{eqnarray}  
where $\langle b_{abc}\rangle^{1}$ is estimated using the regression
coefficient of a linear fit to the mean values of 1,000 simulations
with different $f_{\rm NL}$ values. For $C^{-1}$ we assume that it is
independent of $f_{\rm NL}$, which is a good approximation in the
limit of weak non-Gaussianity, and it is thus estimated with Gaussian
simulations ($\sim 25,000$). The term $\langle T_aT_b\rangle$ is
estimated with 1,000 Gaussian simulations. 
\subsection{Approximated maximum likelihood estimator with diagonal covariance matrix (AMLED)} 

The estimator proposed by \citet{Bucher2010} used the approximation of
assuming a diagonal covariance matrix. In this case, the estimator
simplifies significantly, since the covariance matrix does not need to
be estimated or inverted, and one obtains
\begin{equation}   
f_{\rm NL}=\sum_{\tiny{abc}}\frac{\langle b_{abc}\rangle^{1}/{\rm var}(b_{abc})b_{abc}^{obs}}{\displaystyle\sum_{\tiny{def}}(\langle b_{def}\rangle^{1})^2/{\rm var}(b_{def})}
\label{chi2_fnl}
\end{equation}  
where $var(b_{abc})$ is the variance of the binned bispectrum
components, which is computed with simulations. Besides its
computational efficiency, another advantage of this estimator is that
can be obtained analytically \citep[see][for details]{Bucher2010}.

Strictly speaking, this estimator is optimal only for a full-sky CMB
experiment with isotropic noise (although it has been shown to work
well also in presence of a reduced symmetric mask). Under realistic
conditions, a linear term of a similar form to that used above needs
to be added, such that
\begin{eqnarray} 
f_{\rm NL}=\sum_{\tiny{abc}}\frac{\langle b_{abc}\rangle^{1}/{\rm var}(b_{abc})}{\displaystyle\sum_{\tiny{def}}(\langle b_{def}\rangle^{1})^2/{\rm var}(b_{def})}
\Big(\frac{1}{4\pi N_{pix}}\sum_i^{N_{pix}}T_{a,i}T_{b,i}T_{c,i}^{obs}\\
\nonumber-\langle T_{a,i}T_{b,i}\rangle T_{c,i}-\langle T_{a,i}T_{c,i}\rangle T_{b,i}-\langle T_{b,i}T_{c,i}\rangle T_{a,i} \Big)\;
\label{eq:linterm_diag}
\end{eqnarray}   
As with the previous estimator, 1,000 simulations were used for the
model estimation and another 1,000 to obtain ${\rm var}(b_{abc})$. This
implies a reduction by a factor $>10$ in the number of simulations
required with respect to the AMLE.
    
\subsection{Neural network estimator (NNE)}

\label{subsec:nne}
The architecture of our 3-layer neural network is defined by three
parameters: the number of input, output and hidden nodes. The first
two are determined by the problem at hand; in this case the dimension
of the input vector depends on the number of bins chosen and there is
a single output.

Although the {\sc MemSys} algorithm provides routines to determine the
optimal value of the number of hidden nodes using the Bayesian
evidence \citep{Gull1999}, in this application $n_{\rm hid}$ is
determined empirically by measuring the accuracy of the trained networks on
an independent testing set. In this application, we have found that in
fact the optimal architecture contains no hidden nodes, resulting in
what is effectively a linear mapping between input and output. This is not
surprising, since we are effectively `asking' the network to learn the
mean value and dispersion of the binned bispectrum components for each
$f_{\rm NL}$; since the expectation value is linearly dependent on the
$f_{\rm NL}$, this network architecture trivially satisfies this
requirement. Indeed, networks of this sort provide a simple way of
obtaining the (pseudo)inverse of any matrix.

Then, for zero hidden nodes, the single network output is just a
linear function of the inputs. Once the network parameters
($\vec{w},\theta$) are found during the training process, the
estimator for $f_{\rm NL}$ is thus given by:
\begin{equation}  
f_{\rm NL}=\sum_{abc}w_{abc}b_{abc}+\theta\;.\label{eq:fnl_nn}
\end{equation}  
As with the previous estimators the network is also  trained 
including the linear term, in which case
\begin{gather} 
\label{caca}
f_{\rm NL}=\sum_{abc}w_{abc}\Big(\frac{1}{4\pi N_{pix}}\sum_i^{N_{pix}}T_{a,i}T_{b,i}T_{c,i}-\\\nonumber\langle T_{a,i}T_{b,i}\rangle T_{c,i}-\langle T_{a,i}T_{c,i}\rangle T_{b,i}-\langle T_{b,i}T_{c,i}\rangle T_{a,i} \Big)+\theta\;.
\end{gather}   
Comparing with the AMLE estimator, we can see that it is equivalent to 
a neural network with parameters
 \begin{gather} 
w_{def}\mapsto\sum_{\tiny{abc}}\frac{\langle b_{abc}\rangle^{1}C^{-1}_{abc,def}}{\displaystyle\sum_{\tiny{abc,def}}\langle b_{abc}\rangle^{1}C^{-1}_{abc,def}\langle b_{def}\rangle^{1}}\;.\label{eq:weightsNN} \\
\theta \mapsto 0
\end{gather} 
 If this were the optimal linear combination to estimate $f_{\rm
    NL}$, the neural network would find the same result as the AMLE
  but avoiding all the expensive calculations required in the direct
  computation of this estimator (provided that we have chosen a linear
  combination for the NNE). Conversely, if that combination were
  not optimal, the network should be able to find different, more
  optimal, weights. For instance, for the AMLE to be optimal, the
  considered statistics should follow a Gaussian distribution, whereas
the NNE does not make any assumptions about the intrinsic distribution of
the inputs. Therefore, the neural network is expected to perform
better when working with non-Gaussian statistics. In addition, the neural network does not require to assume
that the covarinace matrix is independent of $f_{\rm NL}$. Even if
this approximation works well for the current application, it may not
always be the case, which would significantly complicate the
calculation of the AMLE.  In such cases the NNE would represent a clear advantage over the
$\chi^2$ minimization. Finally, we would also like to point out that,
although for the current application a linear combination was found to
be the best choice for the NNE, in a general case, this estimator is
not restricted to a linear combination of the inputs, which can be
useful in other problems.

\section{Implementation}
\label{sec:method}
 In this section the non-Gaussian simulations used for the
  analyses as well as some technical details required for the
  implementation of the estimators are described.

\subsection{Non-Gaussian simulations}
\label{subsec:simus}
Two different sets of non-Gaussian realizations are used. A set
generated with the map-making method proposed in \citet{Fergusson2010}
and described also in \citet{Curto2011}, and a set of publicly
available
realisations\footnote{http://planck.mpa-garching.mpg.de/cmb/fnl-simulations/}
generated by \citet{Elsner2009}.
In the first method, the non-Gaussian part of the map ($a_{\ell
  m}^{NG}$) is taken directly from the theoretical bispectrum, while the second
algorithm starts from the primordial curvature fluctuations and is therefore
more precise.

The two different sets are used for the following reasons. Having a
large number of independent realizations is necessary to train the
network, as well as to test its performance with the number of
training data. Since the first set is faster to produce, 30,000
independent realisations were generated as in Curto et al. (2011).  In
the analysis with the SMHW of \citet{Curto2011}, they found the
dispersion on $f_{\rm NL}$ to be slightly larger than using the
simulations of set 2. In \citet{Curto2011}, constraints on $f_{\rm NL}$
are obtained with both sets finding a discrepancy of 5\%. We find
similar deviations for the binned bispectrum.  This is observed if the
average bispectrum at the numerator in (\ref{eq:optimal_sigma}) is
computed with simulations with both sets. Then, as the model of set 1
is given by an approximation, the minimum dispersion of the parameter obtained with realisations is
slightly larger than using the analytical dispersion in
eq.~\ref{eq:optimal_sigma}. Conversely, using realisations of set 2 we
find a closer value to the analytically computed lower bound.

Hence, after proving that the NNE converges with few thousand
realisations for the best performing form of the estimator, the second
set is used for the final results. This is convenient to be able to
compare our results with the Fisher dispersion of
(\ref{eq:optimal_sigma}), and with the ones obtained with the
optimal estimator \citep{Komatsu2011}, where simulations equivalent to
the ones of set 2 are used. 

The Gaussian and non-Gaussian harmonic coefficients of the CMB
realisations, $a_{lm}^{\rm NG}$ and $a_{lm}^{\rm G}$, either generated
from set 1 or set 2, are combined to obtain the non-Gaussian realisation with different values of $f_{\rm NL}$:
 \begin{equation}
a_{lm}=a_{lm}^{\rm G}+f_{\rm NL} a_{lm}^{\rm NG}\;.
\end{equation}
Noise-weighted V+W band WMAP-7yr realizations were
then constructed as explained in \cite{Curto2009} and
\cite{Casaponsa2011a}, and the KQ75 mask was then applied, which
covers roughly $29\%$ of the sky. 

\subsection{Binning scheme}
\label{subsec:binning}

One is free to choose the number and size of the bins in $\ell$-space
for the binned bispectrum.  \citet{Bucher2010} found that for
$\ell_{max}=$2000 and 64 bins the results obtained were 99.3\% of the
optimal value.  For an application to WMAP, one has $\ell_{max}=$1024,
so the corresponding number of bins is 32. We have tested the performance of the estimators with different number of bins and find that for $n_{bin}=28$ the results have converged. Therefore, the following results use
this number of bins, which also provides a modest saving in
computation with respect to 32 bins. Conversely to the exhaustive choosing of the binning scheme done in Bucher et al. (2010) estimator, here we simply use logarithmic
bins. The logarithmic scale is chosen by imposing the condition that all bins
have at least one $\ell$.  

The binned bispectrum components are computed from combinations of
three binned maps $T_{a}T_{b}T_{c}=\sum_{\ell_1\in I_a}\sum_{\ell_2\in
  I_b}\sum_{\ell_3\in I_c}T_{\ell_1}T_{\ell_ 2}T_{\ell_3}$. It can be
noticed that some of the combinations $\ell_1\ell_2\ell_3$ might not
satisfy the triangle condition ($\ell_3-\ell_2\leq\ell_1\leq\ell_2+\ell_3$). To avoid as far as possible those
undesirable combinations, we discard the binned bispectrum
components where all the contained $\ell$ combinations do not meet
the triangle condition. For that reason the components used are the
ones that hold the following condition:
\[\ell^{min}_{I_c}-\ell^{max}_{I_b}\le\ell^{max}_{I_a}\le\ell^{max}_{I_c}+\ell^{max}_{I_b}\;,\] where $\ell^{min}_{I_n}$ and $\ell^{max}_{I_n}$ are the minimum and maximum value of $\ell$ of the bin $I_{n}$. 
Then, the binned bispectrum for $n_{bin}=28$ consists of 1077 components, whereas the full bispectrum would have $\sim10^8$ components.  


\subsection{Inpainting}
\label{subsec:Inpainting}
Several inpainting methods have been developed for general imaging
reconstruction \citep[see e.g. the review by][]{BertalmioWeb}.  The
goal of these methods is to restore missing or damaged regions of an
image to recover the original signal as far as possible. For CMB map
reconstruction, the ideal inpainting method would lead to a restored
map preserving the statistical properties of the unmasked
map.

Different approaches have been used to reduce the discontinuities
generated by the mask edges in CMB maps, since they introduce
undesirable correlations among the binned bispectrum components. As
the intention here is to reduce this impact, rather than reconstruct
the full map, we use a simple iterative process that averages over the
direct neighbours of the masked pixels, and is based on the work of
\citet{Oliveira2001}. 

\begin{figure*}
\centering
\includegraphics[scale=0.5]{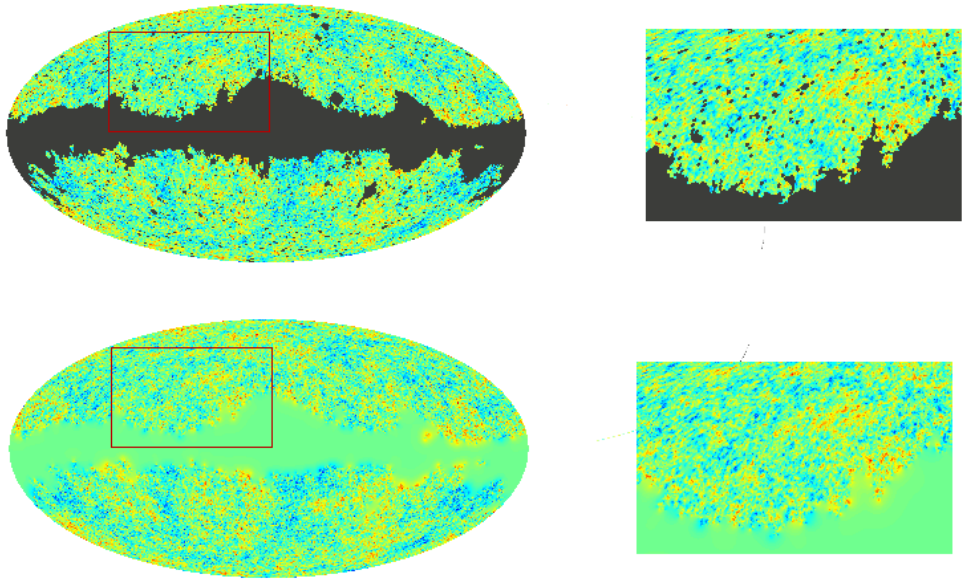}
\caption{Inpainting effect shown in the masked WMAP-7yr map. On the top the initial temperature map with the mask in grey and an amplified region are presented and on the bottom, the same map and region are given after inpainting.}\label{fig:inpainting}
\end{figure*}

One begins with the map $T(\vec{x})$ and the binary mask
$M(\vec{x})$. Then each pixel of the masked map $T'=T\times M$ with
value zero is substituted by the average of its immediate neighbours,
whether masked or not, using the {\sc HEALPix} subroutine {\it
  neighbours}. The process is repeated 1,000 times, leaving the masked
point sources completely inpainted and smoothing the edges of the
galactic mask. The results of this process are illustrated in
Fig.~\ref{fig:inpainting}. We find that, in this case, the technique
is more effective than simply using an apodized mask.

\subsection{Neural network training process}
\label{subsubsec:training}
To train our $f_{\rm NL}$ network we provide it with an ensemble of
training data $\mathcal{D} = \{\mathbf{x}^{(n)},t^{(n)}\}$. The
$n^{\rm th}$ input vector $\mathbf{x}^{(n)}$ contains the binned
bispectrum components, explained in
Section~\ref{sec:binned_bispectrum}, of the $i^{\rm th}$ simulated CMB
map.  The output target is the corresponding $f_{\rm NL}$ value of the
$i^{\rm th}$ CMB simulation.  Thus, for $n_{bin}=28$ the input vector
has 1077 components, and the target vector $t^{(n)}$ for the network
consists of only one component. From the training set, 20 per cent of the realisations
are reserved for the validation process.

The network weights are computed during the training procedure, which
in this case requires only a few seconds. The performance of the
network is validated during the training process using an independent set of testing data.
\begin{figure}
\begin{center}
\includegraphics[scale=0.6]{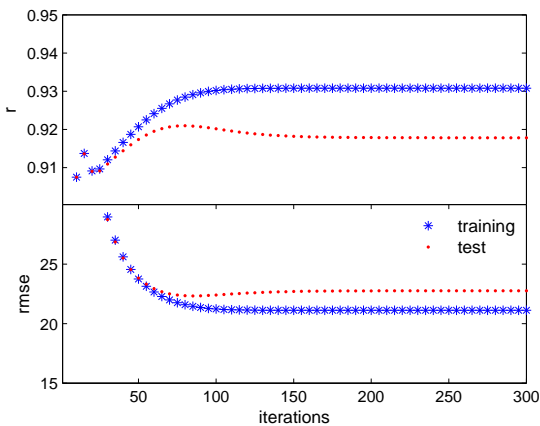}
\end{center}
\caption{In the top panel the Pearson correlation coefficient between
  true $f_{\rm NL}$ value and the network estimator $\hat{f}_{\rm NL}$
  for case 3 of table~\ref{tab:res} versus the number of iterations. Bottom panel is for the root mean
  squared error of $f_{\rm NL}$ at each iteration. Asterisks denote
  training data and dots denote validation
  data.}\label{fig:train_test}
\end{figure}
Figure \ref{fig:train_test} illustrates the training evolution for the
regression network with $n_{\rm hid}=0$ and $n_{\rm data}=10,000$. In
the top panel we plot the correlation coefficient between the target
and the network outputs on the training set and the test set. We see
that a divergence occurs around 60 iterations of the {\sc MemSys}
optimiser due to over-fitting. The same behaviour is confirmed if the
root mean squared error is studied (bottom panel). The network parameters use to construct our final network estimator in
(\ref{eq:fnl_nn}) and (\ref{caca}) are the ones that give a
maximum value of the correlation coefficient and a minimum of the root
mean squared error
in the validation data set.


It is worth noting that for training the neural network, we need
  to choose a certain range of $|f_{\rm NL}|$ to generate the
  required simulations. We find that [-220 220] is
a safe interval for training the network, without significantly biasing the results for $|f_{\rm NL}|$ up to 30.


\section{Results}
\label{sec:results}

As a preliminary check, we applied the three estimators to Gaussian
full-sky maps without noise, finding very similar results in all
cases (see table~\ref{tab:ideal}). 
\begin{table} 
\centering
\begin{tabular}{c|c|c|c}\hline\hline
 Estimator &$\sigma_{fh}$ & $\sigma_g$ & $<f_{\rm NL}>^{Gauss}$ \\\hline\hline  
 AMLED  & \multirow{3}{*}{9.7} & 9.7 & -0.2 \\\cline{1-1}\cline{3-4} 
 AMLE &  & 10.3 & -0.3 \\\cline{1-1}\cline{3-4}                            
  NN & & 9.8 & -0.2  \\\hline\hline
\end{tabular}
\caption{Results for noiseless full sky maps of set 1. The first column is for
  the estimator used, second column indicates the expected dispersion
  for $\ell_{max}=1024$ and in the last two columns are shown the
  dispersion and mean value found for 1,000 Gaussian
  maps.}\label{tab:ideal}
\end{table}
In this ideal case, the AMLE should in principle coincide exactly with
the AMLED, but because of the lack of correlations among the binned
bispectrum components the AMLED seems to be more efficient.  This is
probably due to numerical uncertainties that arise in the covariance
matrix estimation. The neural network is found to be nearly as
efficient as the AMLED.

An important difference between the estimators is the total number of
realisations required to converge, which is directly related to the
computational efficiency. For the AMLED, a few hundred realisations
are sufficient to estimate the variance of the binned bispectrum.  For
the AMLE estimator, however, it is necessary to estimate the
covariance matrix, which requires at least 25,000 Gaussian simulations.
For the NNE, a few thousand realisations are required for the training
process to converge. Nonetheless, it is worth noting that the number
of training realisations required by the NNE does vary with the case
being studied. For example, for inpainted maps where neither the
linear term is taken into account nor the mean is subtracted (case
1 of table~\ref{tab:res}), the NNE needs ~10,000 independent
simulations to converge. 
\begin{figure}
\begin{center}
\includegraphics[scale=0.6]{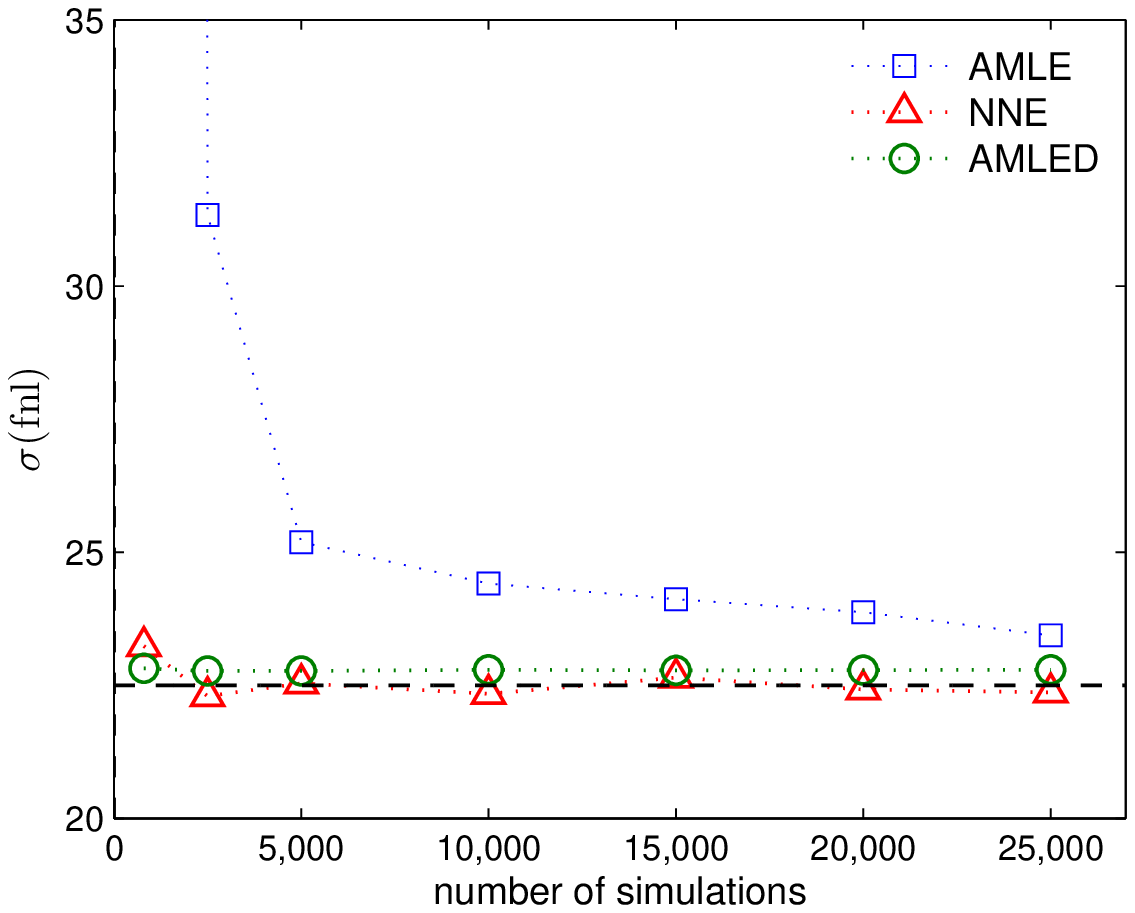}
\includegraphics[scale=0.6]{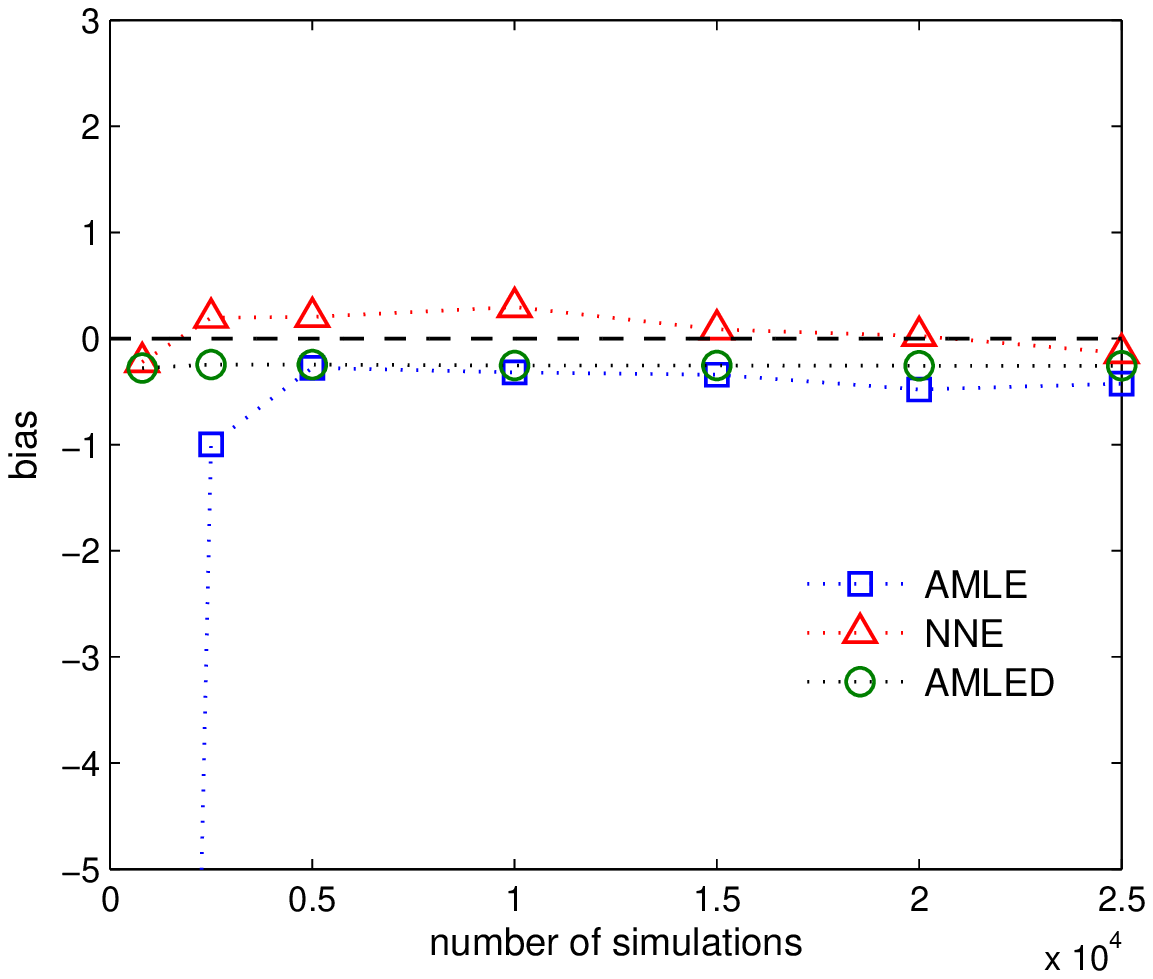}
\end{center}
\caption{Comparison of the efficiency (top) and bias (bottom) of the three estimators
  with respect to the number of simulations used to construct the estimator. For reference, the
  optimal values for the dispersion and bias (dashed black line) are
  also shown. Note that for the NNE, the simulations are used for the
  training process, whereas for the AMLE they are employed to estimate
  the covariance matrix. For the AMLED, they correspond to the number
  of simulations used to obtain the diagonal elements of the
  covariance matrix. }\label{fig:ndata}
 \end{figure}

In applying the three estimators to realistic simulations, based on
WMAP-7yr data, larger differences are observed in the results; these 
are summarised in table~\ref{tab:res}.  We find that
the AMLED estimator reaches values close to the expected dispersion if
and only if the linear term is subtracted and inpainting is
performed. Actually, if the estimator is applied to non-inpainted
maps, the dispersion worsens by a $\sim60\%$. Of course, in the absence
of the linear term, the estimator becomes highly suboptimal, giving
errors of $300\%$. This is not the case for the other two
estimators. We notice that the full covariance matrix $\chi^2$
estimator and the neural network give similar results if instead of
taking into account the linear term, the mean value of the
intermediate maps is subtracted, as is the case for wavelets and
needlets \citep{Donzelli2012}. This is observed in both inpainted and
non-inpainted maps, comparing cases 2 and 3 and 4 and 5 respectively (see
table~\ref{tab:res}). Indeed, these estimators appear more robust,
since the improvement due to the inpainting is small.  In particular,
comparing cases 2 and 4, the NNE estimator without inpainting increases
the dispersion only by 5\% and for the full $\chi^2$ estimator by
$\sim10\%$, while for the AMLED the results are much worse.  Although
similar results are found with the AMLE and the NNE estimators, one
important difference is the number of simulations required to
construct them.  As commented before, 25,000 Gaussian realizations
were used to estimate the covariance matrix in AMLE. As shown in top
panel of Fig.~\ref{fig:ndata}, the NNE requires dramatically fewer
training realisations and also has the advantage that the average
value of the binned bispectrum at $f_{\rm NL}=1$ does not need to be
estimated.  In the same figure, bottom panel, we plot the bias found
for the $f_{\rm NL}$ estimates for 1,000 Gaussian realisations for the three estimators with the number of simulations used.  One sees that the
AMLE requires more realizations than the other two estimators to produce
unbiased results.

\begin{figure}
\begin{center}
\includegraphics[scale=0.6]{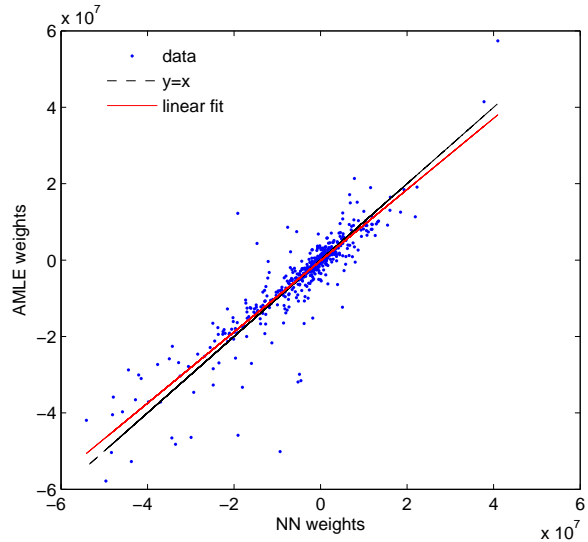}
\end{center}
\caption{Weights for the AMLE estimator involving the covariance matrix and the model, versus the NN weights obtained after the training process. This comparison is made when both estimators have converged presenting a linear fit slope and intercept of $a=91,b=2\times10^5$.}\label{fig:weights}
\end{figure}

All these results indicate that the neural network is a viable short
cut to obtaining the necessary weights to construct the AMLE
estimator. In Fig.~\ref{fig:weights} the weights found for the neural
network are compared to those of the AMLE. Note that the weights
of both estimators are very similar,
validating the relation stated in (\ref{eq:weightsNN}). The contribution of the network parameter $\theta$ is negligible for all cases.

In terms of computational demand, the most efficient estimators are
the NNE and the AMLED, with the number of simulations required
at least 10 times smaller than for the AMLE. Note that for
the AMLED we have used realisations to estimate the average of the
bispectrum at $f_{\rm NL}=1$, therefore the final number of
realisations employed is similar to the ones used for training the NNE. 
\begin{table*}
\centering
\begin{tabular}{c|c|c|c|c|c|c|c}\hline\hline
Casuistry& Inpainting&linear term & mean subs.& Estimator &  $\sigma_g$ & $<f_{\rm NL}>^{Gauss}$&$(\sigma_{fh}-\sigma_g)/\sigma_{fh} (\%)$  \\\hline\hline  

\multirow{3}{*}{1}&\multirow{3}{*}{ Yes }&\multirow{3}{*}{No}&\multirow{3}{*}{No} & AMLED  & 107 & 3 & 300\\\cline{5-8}
&   & &                            & AMLE & 32.7 & -1 & 45\\\cline{5-8}                            
 &  & &                            & NN & 29.7 & -0.3 & 32 \\\hline\hline
\multirow{3}{*}{2}&\multirow{3}{*}{Yes} &\multirow{3}{*}{ Yes} &\multirow{3}{*}{No} & AMLED &  \textbf{22.7} & 0.7 & 0.9\\\cline{5-8}           
& & & & AMLE & {\bf 23.3} & 0.7 & 3.5  \\\cline{5-8}
  &  & & & NN & \textbf{22.4} & 0.7 & 0.4\\\hline\hline
\multirow{3}{*}{3}&\multirow{3}{*}{Yes} &\multirow{3}{*}{ No} &\multirow{3}{*}{ Yes} & AMLED & 31.5 & 0.7& 40 \\\cline{5-8}
 & & & & AMLE& 24.0 & 0.7 & 6.7\\\cline{5-8}     
   & & & & NN & 23.1  & 0.5 & 2.7 \\\hline\hline
\multirow{3}{*}{4}&\multirow{3}{*}{No}&\multirow{3}{*}{Yes} &\multirow{3}{*}{ No} & AMLED & 35.9 & -0.3 & 60  \\\cline{5-8}
& & & & AMLE & 24.3 & 0.1& 9.3  \\\cline{5-8}                            
&  & & & NN & 23.6  & 0.6 & 4.8 \\\hline\hline
\multirow{3}{*}{5}&\multirow{3}{*}{No} &\multirow{3}{*}{ No} &\multirow{3}{*}{ Yes} & AMLED & 37.0 & 1.5  &64\\\cline{5-8}
& & & & AMLE & 24.6 & -0.4& 8.0 \\\cline{5-8}                            
& & & & NN & 23.6 & 0.4 & 4.8\\\hline\hline
 \end{tabular}

\caption{Comparison of results depending on the estimator. The columns are the characteristic of the estimator, if an inpainting of the simulations is made, if the linear term is added and if the mean was subtracted on the binned intermediate maps. Next columns are $\sigma(f_{NL})$ and $\langle f_{\rm NL}\rangle$ for 1,000 Gaussian simulations. Finally the relative error related to the minimum expected dispersion is shown in the last column.}	\label{tab:res}
\label{tabla}
 \end{table*}

For all three estimators, the best results are obtained when the map
is inpainted and the linear term is subtracted (see case 3 of
table~\ref{tab:res}, indicated in bold face). For this optimal case, we compute $\langle b_{abc}\rangle^{1}$ with
1,000 simulations of set 2 \citep{Elsner2009}, to compare it with the
expected dispersion for a WMAP-7yr characteristics, computed as in
(\ref{eq:optimal_sigma}). The neural network is now trained with this
set of $a_{\ell m}$. As we have seen, the NNE typically requires 2,500
independent training realisations to converge. Since only 1,000 are
available, we therefore generated 10,000 simulations using the same
set of $a_{\ell m}$ rotating them and adding different noise contributions. This
procedure was used in \citet{Casaponsa2011b} and was found to be
useful when only a small number of realisations is available.  

In table~\ref{tab:results_wandelt} the final results for all of the
estimators are shown. The values for WMAP-7yr data are without 
point sources correction, which is given in the last column of the
same table. The unresolved point sources contribution to $f_{\rm NL}$
is obtained using the same procedure as in \citet{Curto2009} and
\citet{Casaponsa2011a}.  As expected, by looking at the preliminary
results, the tightest constraints are given by the NNE and AMLED
estimators. For comparison, the WMAP-7yr map $f_{\rm NL}$ estimate
with the optimal estimator obtained by \citet{Komatsu2011} is 42,
without the point sources correction. Note that the closest value is
given by the NNE. The constraints for $f_{\rm NL}$ with the point
source contribution taken into account at 95\% confidence level are
$-3<f_{\rm NL}<83$ to be compare with $-2<f_{\rm NL}<82$ given by the optimal estimator.
\begin{table*}
\centering
\begin{tabular}{c|c|c|c|c|c}\hline\hline
Estimator& $\sigma_{fh}$& $\sigma_g$ & $<f_{\rm NL}>^{Gauss}$ & $f_{\rm NL}^{map}$&$\Delta f_{\rm NL}$\\\hline\hline  

AMLED & \multirow{3}*{21.3} & 21.7 & -0.2 & 33.4 & 3$\pm$2\\\cline{1-1}\cline{3-6}           
AMLE &   & 22.4 & -0.1 & 39.8 & 3$\pm$2\\\cline{1-1}\cline{3-6}
NN & & 21.4 & 0.5 &  44.2 & 4$\pm$2\\\hline\hline
 \end{tabular}

\caption{Results for inpainted Gaussian realizations. Model estimated and neural network trained using Elsner \& Wandelt simulations (set 2). The columns from left to right are: the estimator used, the Fisher $\sigma$ computed from eq.~\ref{eq:optimal_sigma}, the dispersion and mean value of $\hat{f}_{\rm NL}$ for 1,000 Gaussian simulations. Followed by the $f_{\rm NL}$ value found for WMAP-7yr data and the contribution expected by the unresolved point sources ($\Delta f_{\rm NL}$).}	\label{tab:results_wandelt}
\label{tabla}
\end{table*}

\section{Conclusions}
\label{sec:conclusions}

We have trained a regression network with the binned bispectrum
components of non-Gaussian realizations in order to obtain 
constraints on the local non-linear coupling parameter $f_{\rm
  NL}$. We have compared the results with those obtained with a
maximum-likelihood estimator, using either a diagonal or a full covariance
matrix.  We also studied the effect of the addition of the linear
term, mean subtraction and the use of inpainting.  

We find that the three estimators become close to optimal if the
linear term is subtracted and inpainting is performed. We find that
the linear term is absolutely necessary if a diagonal covariance
matrix is used. However, its effect is very small if the full
covariance matrix or the neural network is used and the mean is
subtracted from the binned maps, as found for wavelets and needlets in
\citet{Donzelli2012} and \citet{Curto2012}. In that sense, the choice
of the estimator depends on the difficulty of computing the linear
term.  Although the best results for all estimators are obtained when
inpainted maps are used, the largest effect of this technique is seen
in the AMLED estimator, with the other two being less affected by the
presence of a mask.  Thus, the most robust tools are the AMLE and the
NNE estimators, with the NNE displaying a clear computational
advantage, since the covariance matrix does not need to be estimated
or inverted; this reduces significantly the number of simulations
required.  Another advantage of the neural network estimator arises
from the fact that for $\chi^2$ minimization the dependence of the
covariance matrix on $f_{\rm NL}$ makes a full solution
computationally hard, if not unfeasible, for certain
problems. Conversely, the NNE bypasses such calculations, thereby
simplifying the analysis.
 
 
We conclude that the most efficient tools are the neural network
regression estimator and the AMLED estimator. The latter would be the
choice if a small set of non-Gaussian simulations is available
($\sim$1,000), or analytical models are preferred. However, the AMLED
depends on a specific pre-processing of the data. Neural networks
give almost optimal results, without the use of inpainting,
thereby avoiding the need to alter the data.  

Finally, the constraints for WMAP-7yr data, with the unresolved point
sources contribution included, at 95\% confidence level would be
$-3<f_{\rm NL}<83$.  These results are compatible with $f_{\rm NL}=$0,
as found in \citet{Komatsu2011,Curto2011,Bennett2012}. Note that we
have used foreground reduced maps, and the foregrounds have not been
marginalised over in this analysis.

We note that neural networks would be a useful method to estimate jointly other
forms of non-Gaussianity, such as those where the number of outputs were set to a
number of different $f_{\rm NL}$ shapes (e.g. local, equilateral,
orthogonal), but this is left for future work.

\section*{acknowledgments}
The authors thank Bartjan van Tent for useful discussions. We acknowledge partial financial support from the Spanish Ministerio
de Ciencia e Innovaci\'on project AYA2007-68058-C03-02 and
AYA2010-21766-C03-01. We acknowledge partial
financial support from the Spanish Minsterio de Econom\'{i}a
y Competitividad AYA2010-21766-C03-01 and Consolider-
Ingenio 2010 CSD2010-00064 projects. 
B. Casaponsa thanks the Spanish Ministerio de
Ciencia e Innovaci\'on for a pre-doctoral fellowship. The
authors acknowledge the computer resources, technical expertise and
assistance provided by the Spanish Supercomputing Network (RES) node
at Universidad de Cantabria. We acknowledge the use of Legacy Archive
for Microwave Background Data Analysis (LAMBDA). Support for it is
provided by the NASA Office of Space Science. The {\sc HealPix} package was
used throughout the data analysis \citep{Gorski2005}.

\bibliographystyle{mn2e}
\bibliography{citas}
\end{document}